\documentclass[letterpaper,english,onecolumn]{IEEEtran}
\usepackage[T1]{fontenc}
\usepackage[latin9]{inputenc}
\usepackage{refstyle}
\usepackage{amsmath}
\usepackage{amsthm}
\usepackage{amssymb}

\makeatletter

\AtBeginDocument{\providecommand\eqref[1]{\ref{eq:#1}}}
\pdfpageheight\paperheight
\pdfpagewidth\paperwidth

\RS@ifundefined{subsecref}
  {\newref{subsec}{name = \RSsectxt}}
  {}
\RS@ifundefined{thmref}
  {\def\RSthmtxt{theorem~}\newref{thm}{name = \RSthmtxt}}
  {}
\RS@ifundefined{lemref}
  {\def\RSlemtxt{lemma~}\newref{lem}{name = \RSlemtxt}}
  {}

\theoremstyle{plain}
\newtheorem{thm}{\protect\theoremname}
\theoremstyle{plain}
\newtheorem{prop}[thm]{\protect\propositionname}
\theoremstyle{plain}
\newtheorem{lem}[thm]{\protect\lemmaname}
\theoremstyle{remark}
\newtheorem{rem}[thm]{\protect\remarkname}

\makeatother

\usepackage{babel}
\providecommand{\lemmaname}{Lemma}
\providecommand{\propositionname}{Proposition}
\providecommand{\remarkname}{Remark}
\providecommand{\theoremname}{Theorem}

\begin{document}
\title{Properties of a Generalized Divergence Related to Tsallis Relative
Entropy}
\author{Rui F.\ Vigelis\thanks{R. F. Vigelis is with Computer Engineering, Campus Sobral, Federal University of Cear\'{a}, 62.010-560, Sobral-CE, Brazil, e-mail:rfvigelis@ufc.br}, Luiza H.F.\ de Andrade\thanks{L. H.F. Andrade is with Department of Natural Sciences, Mathematics and Statistics, Federal Rural University of the Semi-arid Region, 59.625-900, Mossor\'{o}-RN, Brazil, e-mail:luizafelix@ufersa.edu.br}, Charles C.\ Cavalcante, ~\IEEEmembership{Senior Member,~IEEE,}\thanks{C. C. Cavalcante is with Department of Teleinformatics Engineering, Federal University of Cear\'{a}, 60020-181, Fortaleza-CE, Brazil, e-mail:charles@ufc.br} \thanks{Copyright (c) 2017 IEEE. Personal use of this material is permitted.  However, permission to use this material for any other purposes must be obtained from the IEEE by sending a request to pubs-permissions@ieee.org.}}%

\maketitle

\begin{abstract}
In this paper, we investigate the partition inequality, joint convexity,
and Pinsker's inequality, for a divergence that generalizes the Tsallis
Relative Entropy and Kullback\textendash Leibler divergence. The generalized
divergence is defined in terms of a deformed exponential function,
which replaces the Tsallis $q$-exponential. We also constructed a
family of probability distributions related to the generalized divergence.
We found necessary and sufficient conditions for the partition inequality
to be satisfied. A sufficient condition for the joint convexity was
established. We proved that the generalized divergence satisfies the
partition inequality, and is jointly convex, if, and only if, it coincides
with the Tsallis relative entropy. As an application of partition
inequality, a criterion for the Pinsker's inequality was found.
\end{abstract}


\section{Introduction}

Statistical divergences play an essential role in Information Theory
\cite{Cover2006}. Divergence can be interpreted as a measure of dissimilarity
between two probability distributions. Applications that use it span
from areas such as communications to econometric and other physical
systems \cite{Cover2006}. Entropy can be derived from the notion
of divergence. Numerous definitions of divergence can be found in
the literature. The interest in different statistical divergences
is motivated by applications related to optimization and statistical
learning, since more flexible functions and expressions may be suitable
for larger classes of data and signals, leading to more efficient
information recovery methods \cite{Hastie2009,Principe2010,Konishi2008}.
The divergence usefulness depends on its properties, such as non negativity,
monotonicity and joint convexity, among others.

The counterpart of Shannon entropy is the well-known Kullback\textendash Leibler
(KL) divergence \cite{Kullback1951}, denoted by $D_{\mathrm{KL}}(\cdot\mathbin{||}\cdot)$,
which is extensively used in Information Theory. Tsallis relative
entropy $D_{q}(\cdot\mathbin{||}\cdot)$, which generalizes KL divergence,
is defined in terms of the $q$-logarithm \cite{Borland1998,Borland1999}.
Both KL divergence and Tsallis relative entropy satisfy some important
properties, such as non negativity, joint convexity, and Pinsker's
inequality \cite{Erven2014,Furuichi2004,Furuichi2005}. A \textit{generalized
divergence} $D_{\varphi}(\cdot\mathbin{||}\cdot)$ can be defined
in terms of a deformed exponential function $\varphi$, which plays
the role of $q$-logarithm in Tsallis relative entropy. The generalized
divergence appeared before in the literature, as a specific case in
a broader class of divergences. Zhang in \cite{Zhang2004} introduced
a divergence denoted by $D_{f,\rho}^{(\alpha)}(\cdot\mathbin{||}\cdot)$,
where $\alpha\in[-1,1]$, and$f$ and $\rho$ are functions. The generalized
divergence corresponds to Zhang's divergence with $\alpha=-1$, and
$\rho=f^{-1}=\varphi^{-1}$ for a deformed exponential function $\varphi$.
In \cite{Broniatowski2019}, another class of divergences was investigated.
The divergences $D_{\beta}^{c}(\cdot\mathbin{||}\cdot)$ in this class
are given in terms of parameters $\beta=(\phi,M_{1},M_{2},M_{3},\lambda)$.
Expression (1) in \cite{Broniatowski2019}, which defines $D_{\beta}^{c}(\cdot\mathbin{||}\cdot)$,
reduces to the generalized divergence, with $\phi=-\varphi^{-1}$,
$M_{1}=1$, $M_{2}=1$, $M_{3}=(\varphi^{-1})'(q)$, and $\lambda=\lambda_{\#}$
is the counting measure.

In \cite{Zhang2004,Broniatowski2019}, the proposed divergences were
investigated from a geometric and minimization perspectives. Some
properties, which are useful in Information Theory, have not been
analyzed for these divergences. In this work, we investigate the partition
inequality, joint convexity, and Pinsker's inequality. We also consider
the family of probability distributions associated with the generalized
divergence $D_{\varphi}(\cdot\mathbin{||}\cdot)$. We showed necessary
and sufficient conditions for the generalized divergence to satisfy
the partition inequality. A sufficient condition for the joint convexity
of $D_{\varphi}(\cdot\mathbin{||}\cdot)$ was found. We proved that
$D_{\varphi}(\cdot\mathbin{||}\cdot)$ satisfies the partition inequality,
and is jointly convex, if, and only if, it coincides with the Tsallis
relative entropy $D_{q}(\cdot\mathbin{||}\cdot)$. Ours results for
Pinsker's inequality are in accordance to previous works \cite{Gilardoni2010,Sason2016}.

The rest of paper is organized as follows. In Section~\ref{sec:defs}
we provide the definition of generalized divergence. Section~\ref{sec:families}
is devoted to the construction of a family of probability distributions.
Properties of the generalized divergence are studied in Section~\ref{sec:properties}.
Finally, conclusions and perspectives are stated in Section~\ref{sec:conclusions}.

\section{Generalized divergence}

The generalized divergence is defined in terms of a deformed exponential
function $\varphi(\cdot)$. Writing the KL divergence or Tsallis relative
entropy in appropriate form, we can obtain the generalized divergence
by replacing $\ln(\cdot)$ or $\ln_{q}(\cdot)$ by the inverse of
a deformed exponential $\varphi^{-1}(\cdot)$. We also provide a construction
of a family of probability distributions related the generalized divergence.

\subsection{\label{sec:defs}Definitions}

For simplicity we denote the set of all probability distributions
on $I_{n}=\{1,\dots,n\}$ by
\[
\Delta_{n}=\biggl\{(p_{1},\dots,p_{n}):\sum_{i=1}^{n}p_{i}=1\text{ and }p_{i}\geq0\text{ for all }i\biggr\}.
\]
The generalized divergence is defined for probability distribution
in the interior of $\Delta_{n}$, which is denoted by $\Delta_{n}^{\circ}$.
A probability distribution $\boldsymbol{p}=(p_{i})$ belongs to $\Delta_{n}^{\circ}$
if and only if $p_{i}>0$ for each $i$,

A \textit{deformed exponential function }is a convex function $\varphi\colon\mathbb{R}\rightarrow[0,\infty)$
such that $\lim_{u\rightarrow-\infty}\varphi(u)=0$ and $\lim_{u\rightarrow\infty}\varphi(u)=\infty$.
It is easy to verify that the ordinary exponential and Tsallis $q$-exponential
are deformed exponential functions. The \textit{Tsallis $q$-exponential}
$\exp_{q}\colon\mathbb{R}\rightarrow[0,\infty)$ is given by
\[
\exp_{q}(x)=\begin{cases}
[1+(1-q)x]_{+}^{1/(1-q)}, & \text{if }q\in(0,1],\\
\exp(x), & \text{if }q=1,
\end{cases}
\]
where $[x]_{+}=x$ for $x\geq0$, and $=0$ otherwise. The\textit{
Tsallis $q$-logarithm} $\ln_{q}\colon(0,\infty)\rightarrow\mathbb{R}$
is defined as the inverse of $\exp_{q}(\cdot)$, which is given by
$\ln_{q}(x)=\frac{1}{1-q}(x^{1-q}-1)$ if $q\in(0,1]$.

Fixed a deformed exponential function $\varphi\colon\mathbb{R}\rightarrow[0,\infty)$,
the \textit{generalized divergence} (or \textit{generalized relative
entropy}) between two probability distributions $\boldsymbol{p}=(p_{i})$
and $\boldsymbol{q}=(q_{i})$ in $\Delta_{n}^{\circ}$ is defined
as
\begin{equation}
D_{\varphi}(\boldsymbol{p}\mathbin{||}\boldsymbol{q})=\sum_{i=1}^{n}\frac{\varphi^{-1}(p_{i})-\varphi^{-1}(q_{i})}{(\varphi^{-1})'(p_{i})}.\label{eq:relative_entropy_generalized}
\end{equation}
Clearly, expression (\ref{eq:relative_entropy_generalized}) reduces
to the KL divergence $D_{\mathrm{KL}}(\boldsymbol{p}\mathbin{||}\boldsymbol{q})=-\sum_{i=1}^{n}p_{i}\ln\bigl(\frac{q_{i}}{p_{i}}\bigr)$
if $\varphi$ is the exponential function. Tsallis relative entropy
in its standard form is given by $D_{q}(\boldsymbol{p}\mathbin{\Vert}\boldsymbol{q})=-\sum_{i=1}^{n}p_{i}\ln_{q}\bigl(\frac{q_{i}}{p_{i}}\bigr)$.
The equality
\[
-p_{i}\dfrac{(q_{i}/p_{i})^{1-q}-1}{1-q}=\frac{1}{p_{i}^{q}}\Bigl(\dfrac{p_{i}^{1-q}-1}{1-q}-\dfrac{q_{i}^{1-q}-1}{1-q}\Bigr)
\]
shows that $D_{q}(\cdot\mathbin{\Vert}\cdot)$ can be written as in
(\ref{eq:relative_entropy_generalized}) if $\varphi$ is the Tsallis
$q$-exponential.

The non-negativity of $D_{\varphi}(\cdot\mathbin{\Vert}\cdot)$ is
a consequence of the concavity of $\varphi^{-1}(\cdot)$. Because
$\varphi^{-1}(\cdot)$ is concave, it follows that
\begin{equation}
(y-x)(\varphi^{-1})'(y)\leq\varphi^{-1}(y)-\varphi^{-1}(x),\qquad\text{for all }x,y>0.\label{eq:phi_inv_conseq}
\end{equation}
Using this inequality with $y=p_{i}$ and $x=q_{i}$, we can write
\[
D_{\varphi}(\boldsymbol{p}\mathbin{\Vert}\boldsymbol{q})=\sum_{i=1}^{n}\frac{\varphi^{-1}(p_{i})-\varphi^{-1}(q_{i})}{(\varphi^{-1})'(p_{i})}\geq\sum_{i=1}^{n}(p_{i}-q_{i})=0.
\]
Its is clear that $D_{\varphi}(\boldsymbol{p}\mathbin{\Vert}\boldsymbol{q})=0$
if $\boldsymbol{p}=\boldsymbol{q}$. The converse depends on whether
$\varphi^{-1}(x)$ is strictly concave. Indeed, if we suppose that
$\varphi^{-1}(x)$ is strictly concave, then an equality in \eqref{phi_inv_conseq}
is attained if and only if $x=y$. Therefore, when $\varphi^{-1}(x)$
is strictly concave, the equality $D_{\varphi}(\boldsymbol{p}\mathbin{\Vert}\boldsymbol{q})=0$
is satisfied if and only if $\boldsymbol{p}=\boldsymbol{q}$.

In addition to similarities between the generalized divergence, KL
divergence, and Tsallis relative entropy, there exists another motivation
for the choice of expression given as in (\ref{eq:relative_entropy_generalized}).
We can associate with the generalized relative entropy $D_{\varphi}(\cdot\mathbin{\Vert}\cdot)$
a $\varphi$-family of probability distributions, just as the KL divergence
is related to the moment-generating function in a exponential family
of probability distributions.

\subsection{\label{sec:families}Families of probability distributions}

For each probability distribution $\boldsymbol{p}=(p_{i})\in\Delta_{n}^{\circ}$,
we can define a \textit{deformed exponential family} (of probability
distributions) \textit{centered at $\boldsymbol{p}$}. A deformed
exponential family consists of a parameterization for the set $\Delta_{n}^{\circ}$.
We remark that a deformed exponential family depends on the centered
probability distribution $\boldsymbol{p}$. We can associate with
each probability distribution $\boldsymbol{p}\in\Delta_{n}^{\circ}$
a deformed exponential family centered at $\boldsymbol{p}$.

Assume that $\varphi\colon\mathbb{R}\rightarrow[0,\infty)$ is a positive,
deformed exponential function with continuous derivative. Fixed $\boldsymbol{p}=(p_{i})\in\Delta_{n}^{\circ}$,
let $\boldsymbol{c}=(c_{i})$ be a vector such that $p_{i}=\varphi(c_{i})$
for each $i$. We also fix a vector $\boldsymbol{u}_{0}=(u_{0i})$
such that $u_{0i}>0$ for each $i$, and
\begin{equation}
\sum_{i=1}^{n}u_{0i}\varphi'(c_{i})=1.\label{eq:u0_constraint}
\end{equation}
A \textit{deformed exponential family (of probability distributions)
centered at $\boldsymbol{p}$} is a parameterization of $\Delta_{n}^{\circ}$,
which maps each vector $\boldsymbol{u}=(u_{i})$ in the subspace
\[
B_{\boldsymbol{c}}^{\varphi}=\biggl\{(u_{1},\dots,u_{n}):\sum_{i=1}^{n}u_{i}\varphi'(c_{i})=0\biggr\}
\]
to a probability distribution $\boldsymbol{q}=(q_{i})\in\Delta_{n}^{\circ}$
by the expression
\begin{equation}
q_{i}=\varphi(c_{i}+u_{i}-\psi_{\boldsymbol{c}}(\boldsymbol{u})u_{0i}),\label{eq:family_phi}
\end{equation}
where $\psi_{\boldsymbol{c}}\colon B_{\boldsymbol{c}}^{\varphi}\rightarrow[0,\infty)$
is the \textit{normalizing function}, which is introduced so that
(\ref{eq:family_phi}) defines a probability density in $\Delta_{n}^{\circ}$.

The choice for $\boldsymbol{u}\in B_{\boldsymbol{c}}^{\varphi}$ is
not arbitrary. Thanks to this choice, it is possible to find $\psi_{\boldsymbol{c}}(\boldsymbol{u})\geq0$
for which expression (\ref{eq:family_phi}) is a probability density
in $\Delta_{n}^{\circ}$. We will justify this claim. Because $\varphi(\cdot)$
is convex, it follows that
\begin{equation}
y\varphi'(x)\leq\varphi(x+y)-\varphi(x),\qquad\text{for all }x,y\in\mathbb{R}.\label{eq:phi_convex_conseq}
\end{equation}
Using (\ref{eq:phi_convex_conseq}) with $x=c_{i}$ and $y=u_{i}$,
we can write, for any $\boldsymbol{u}\in B_{\boldsymbol{c}}^{\varphi}$,
\[
1=\sum_{i=1}^{n}u_{i}\varphi'(c_{i})+\sum_{i=1}^{n}\varphi(c_{i})\leq\sum_{i=1}^{n}\varphi(c_{i}+u_{i}).
\]
By the definition of $\varphi(\cdot)$, the map
\[
g(\lambda)=\sum_{i=1}^{n}\varphi(c_{i}+u_{i}-\lambda u_{0i})
\]
is continuous, approaches $0$ as $\lambda\rightarrow\infty$, and
tends to $\infty$ as $\alpha\rightarrow\infty$. Since $\varphi(\cdot)$
is strictly increasing, it follows that $g(\cdot)$ is strictly decreasing.
Then we can conclude that there exists a unique $\lambda_{0}=\psi_{\boldsymbol{c}}(\boldsymbol{u})\geq0$
for which $q_{i}=\varphi(c_{i}+u_{i}-\lambda_{0}u_{0i})$ is a probability
distribution in $\Delta_{n}^{\circ}$.

The generalized divergence $D_{\varphi}(\cdot\mathbin{\Vert}\cdot)$
is associated with the deformed exponential family (\ref{eq:family_phi})
by the equality
\begin{equation}
\psi_{\boldsymbol{c}}(\boldsymbol{u})=D_{\varphi}(\boldsymbol{p}\mathbin{\Vert}\boldsymbol{q})=\sum_{i=1}^{n}\frac{\varphi^{-1}(p_{i})-\varphi^{-1}(q_{i})}{(\varphi^{-1})'(p_{i})}.\label{eq:psi_D_phi}
\end{equation}
Using $\sum_{i=1}^{n}u_{i}\varphi'(c_{i})=0$, together with the constraint
(\ref{eq:u0_constraint}), we can write
\begin{equation}
\psi_{\boldsymbol{c}}(\boldsymbol{u})=\sum_{i=1}^{n}(-u_{i}+\psi_{\boldsymbol{c}}(\boldsymbol{u})u_{0i})\varphi'(c_{i}).\label{eq:psi_u_be_replaced}
\end{equation}
It is clear that
\begin{equation}
-u_{i}+\psi_{\boldsymbol{c}}(\boldsymbol{u})u_{0i}=\varphi^{-1}(p_{i})-\varphi^{-1}(q_{i}),\label{eq:replace_A}
\end{equation}
and
\begin{equation}
\varphi'(c_{i})=\frac{1}{(\varphi^{-1})'(p_{i})}.\label{eq:replace_B}
\end{equation}
Inserting (\ref{eq:replace_A}) and (\ref{eq:replace_B}) into (\ref{eq:psi_u_be_replaced}),
we obtain (\ref{eq:psi_D_phi}).

If $\varphi$ is the exponential function, and $u_{0i}=1$, the deformed
exponential family reduces to the well known \textit{exponential family}:
\begin{equation}
q_{i}=\exp(u_{i}-K_{\boldsymbol{p}}(\boldsymbol{u}))\cdot p_{i},\label{eq:family_exponential}
\end{equation}
where $K_{\boldsymbol{p}}(\boldsymbol{u})$ is the \textit{cumulant-generating
function}, which equals the normalizing function $\psi_{\boldsymbol{c}}(\boldsymbol{u})$.

\section{\label{sec:properties}Properties of the generalized divergence}

The KL divergence and Tsallis relative entropy satisfy the partition
inequality, and are jointly convex. They also satisfy Pinsker's inequality.
We will investigate under what conditions these properties hold for
the generalized divergence. Throughout this section we assume that
$(\varphi^{-1})''(x)$ is continuous and $>0$.

\subsection{Partition inequality}

Partition inequality, which is a case of the data processing inequality,
will be used in the proof of Pinsker's inequality. Let $\mathcal{A}=\{A_{1},\dots,A_{k}\}$
be a partition of $I_{n}=\{1,\dots,n\}$, i.e., $\mathcal{A}$ is
a collection of subsets $A_{j}\subseteq I_{n}$ such that $A_{i}\cap A_{j}=\emptyset$,
for $i\neq j$, and $\bigcup_{j=1}^{k}A_{j}=I_{n}$. For any probability
distribution $\boldsymbol{p}=(p_{i})$, we define the probability
distribution $\boldsymbol{p}^{\mathcal{A}}=(p_{j}^{\mathcal{A}})$
as
\[
p_{j}^{\mathcal{A}}=\sum_{i\in A_{j}}p_{i},\qquad\text{for each }j=1,\dots,k.
\]

The next result gives a necessary and sufficient condition for the
partition inequality to be satisfied.

\begin{prop}
\label{prop:ineq_partition} For the divergence $D_{\varphi}(\cdot\mathbin{||}\cdot)$
to satisfy the partition inequality
\begin{equation}
D_{\varphi}(\boldsymbol{p}\mathbin{||}\boldsymbol{q})\geq D_{\varphi}(\boldsymbol{p}^{\mathcal{A}}\mathbin{||}\boldsymbol{q}^{\mathcal{A}}),\label{eq:ineq_partition}
\end{equation}
for all probability distributions $\boldsymbol{p}=(p_{i})$ and $\boldsymbol{q}=(q_{i})$,
and any partition $\mathcal{A}$ of $I_{n}$, it is necessary and
sufficient that the function $g=-\dfrac{(\varphi^{-1})'}{(\varphi^{-1})''}$
be superadditive, i.e., the inequality
\begin{equation}
g(x+y)\geq g(x)+g(y),\label{eq:almost_superadditive}
\end{equation}
be satisfied for all $x,y\in(0,1)$ such that $x+y\in(0,1)$.
\end{prop}

The proof of Proposition~\ref{prop:ineq_partition} requires some
preliminary results which are presented in the sequel.

\begin{lem}
\label{lem:F_alpha_subadditive_G_concave} Fix any $\alpha\in(0,1)$.
The mapping
\[
F_{\alpha}(x,y)=\varphi((1-\alpha)\varphi^{-1}(x)+\alpha\varphi^{-1}(y)),
\]
is superadditive in $(0,1)\times(0,1)$ if, and only if,
\[
G(x,y)=\varphi^{-1}(\varphi(x)+\varphi(y))
\]
is convex in $\{(x,y)\in\mathbb{R}^{2}:\varphi(x)+\varphi(y)\in(0,1)\}$.
\end{lem}
\begin{IEEEproof}
Let $x_{i},y_{i}\in(0,1)$ be such that $x_{1}+x_{2}\in(0,1)$ and
$y_{1}+y_{2}\in(0,1)$. The superadditivity of $F_{\alpha}$ implies
that
\begin{multline}
\varphi((1-\alpha)\varphi^{-1}(x_{1}+x_{2})+\alpha\varphi^{-1}(y_{1}+y_{2}))\\
\geq\varphi((1-\alpha)\varphi^{-1}(x_{1})+\alpha\varphi^{-1}(y_{1}))\\
+\varphi((1-\alpha)\varphi^{-1}(x_{2})+\alpha\varphi^{-1}(y_{2})).\label{eq:F_alpha_ineq}
\end{multline}
Denote $s_{i}=\varphi^{-1}(x_{i})$ and $t_{i}=\varphi^{-1}(y_{i})$
for $i=1,2$. Thus inequality \eqref{F_alpha_ineq} is equivalent
to
\begin{multline*}
(1-\alpha)\varphi^{-1}(\varphi(s_{1})+\varphi(s_{2}))+\alpha\varphi^{-1}(\varphi(t_{1})+\varphi(t_{2}))\\
\geq\varphi^{-1}[\varphi((1-\alpha)s_{1}+\alpha t_{1})+\varphi((1-\alpha)s_{2}+\alpha t_{2})],
\end{multline*}
which shows the desired result.
\end{IEEEproof}

\begin{lem}
\label{lem:G_convex_g_superadditive} The function $G$, as defined
in Lemma~\ref{lem:F_alpha_subadditive_G_concave}, is convex if and
only if $g=-\dfrac{(\varphi^{-1})'}{(\varphi^{-1})''}$ is superadditive
in $(0,1)$.
\end{lem}
\begin{IEEEproof}
For the function $G$ to be convex, it is necessary and sufficient
that its Hessian $H_{G}$ be positive semi-definitive, which is equivalent
to $\operatorname{tr}(H_{G})\geq0$ and $J_{G}=\det(H_{G})\geq0$,
where $\operatorname{tr}(\cdot)$ denotes the trace of a matrix and
$\det(\cdot)$ is the determinant of a matrix (see \cite{Bhatia2007}).
Letting $z=\varphi(x)+\varphi(y)$, we can express
\begin{align}
\frac{\partial^{2}G}{\partial x^{2}}(x,y) & =\varphi''(x)(\varphi^{-1})'(z)+[\varphi'(x)]^{2}(\varphi^{-1})''(z),\label{eq:partial_deriv_G_x}\\
\frac{\partial^{2}G}{\partial y^{2}}(x,y) & =\varphi''(y)(\varphi^{-1})'(z)+[\varphi'(y)]^{2}(\varphi^{-1})''(z),\label{eq:partial_deriv_G_y}
\end{align}
and
\begin{equation}
\frac{\partial^{2}G}{\partial x\partial y}(x,y)=\varphi'(x)\varphi'(y)(\varphi^{-1})''(z).\label{eq:partial_deriv_G_xy}
\end{equation}
If we divide the right-hand side of (\ref{eq:partial_deriv_G_x})
by $-\varphi''(x)(\varphi^{-1})''(z)\geq0$, and we use
\begin{equation}
\frac{[\varphi(x)']^{2}}{\varphi(x)''}=-\frac{(\varphi^{-1})'(\varphi(x))}{(\varphi^{-1})''(\varphi(x))}\label{eq:useful_expression_g}
\end{equation}
into the resulting expression, we obtain
\[
-\frac{(\varphi^{-1})'(z)}{(\varphi^{-1})''(z)}+\frac{(\varphi^{-1})'(\varphi(x))}{(\varphi^{-1})''(\varphi(x))}=g(z)-g(\varphi(x)).
\]
As a result, we conclude that $\partial^{2}G/\partial x^{2}\geq0$
(and similarly $\partial^{2}G/\partial y^{2}\geq0$) if $g$ is superadditive.
Using expressions (\ref{eq:partial_deriv_G_x})\textendash (\ref{eq:partial_deriv_G_xy})
for the partial derivatives of $G$, we find
\begin{multline*}
J_{G}(x,y)=(\varphi^{-1})'(z)(\varphi^{-1})''(z)\varphi''(x)\varphi''(y)\\
\cdot\biggl\{\frac{(\varphi^{-1})'(z)}{(\varphi^{-1})''(z)}+\frac{[\varphi'(y)]^{2}}{\varphi''(y)}+\frac{[\varphi'(x)]^{2}}{\varphi''(x)}\biggr\}.
\end{multline*}
In view of (\ref{eq:useful_expression_g}), it follows that $J_{G}(x,y)\geq0$
is equivalent to $g(z)\geq g(\varphi(x))+g(\varphi(y))$. Thus $G$
is convex if and only if $g$ is superadditive in $(0,1)$.
\end{IEEEproof}

\begin{rem}
Similar versions of these lemmas appeared previously in the literature
(see \cite[sec.~3.16]{Hardy1988} and \cite{Matkowski1990}). The
hypothesis in these versions was weaker, or just one direction was
proved.
\end{rem}

Now, we may proceed to the proof of the main result in this section.

\begin{IEEEproof}[Proof of Proposition~\ref{prop:ineq_partition}]
\textit{ Sufficiency}. Lemmas~\ref{lem:F_alpha_subadditive_G_concave}
and~\ref{lem:G_convex_g_superadditive} imply that
\[
F_{\alpha}(x,y)=\varphi((1-\alpha)\varphi^{-1}(x)+\alpha\varphi^{-1}(y)),
\]
is superadditive in $(0,1)\times(0,1)$, for each $\alpha\in(0,1)$.
Considering $\mathcal{A}=\{A_{1},\dots,A_{k}\}$, we denote $p_{j}^{\mathcal{A}}=\sum_{i\in A_{j}}p_{i}$
and $q_{j}^{\mathcal{A}}=\sum_{i\in A_{j}}q_{i}$. By the superadditivity
of $F_{\alpha}(x,y)$, we can write
\begin{multline}
\frac{1}{1-\alpha}\sum_{i=1}^{n}[p_{i}-F_{\alpha}(q_{i},p_{i})]\\
\geq\frac{1}{1-\alpha}\sum_{j=1}^{k}[p_{j}^{\mathcal{A}}-F_{\alpha}(q_{j}^{\mathcal{A}},p_{j}^{\mathcal{A}})].\label{eq:F_alpha_p_q-1}
\end{multline}
An application of L'H\^{o}pital's rule on the limit below provides
\begin{align*}
\lim_{\alpha\uparrow1}\frac{y-F_{\alpha}(x,y)}{1-\alpha} & =\lim_{\alpha\uparrow1}\frac{y-\varphi((1-\alpha)\varphi^{-1}(x)+\alpha\varphi^{-1}(y))}{1-\alpha}\\
 & =\varphi'(\varphi^{-1}(y))[-\varphi^{-1}(x)+\varphi^{-1}(y)]\\
 & =\frac{\varphi^{-1}(y)-\varphi^{-1}(x)}{(\varphi^{-1})'(y)}.
\end{align*}
Thus, in the limit $\alpha\uparrow1$, expression (\ref{eq:F_alpha_p_q-1})
becomes
\[
D_{\varphi}(\boldsymbol{p}\mathbin{||}\boldsymbol{q})\geq D_{\varphi}(\boldsymbol{p}^{\mathcal{A}}\mathbin{||}\boldsymbol{q}^{\mathcal{A}}),
\]
which is the asserted inequality.

\textit{Necessity}. It is clear that if (\ref{eq:ineq_partition})
holds for all $\boldsymbol{p}=(p_{i})$, $\boldsymbol{q}=(q_{i})$,
and $\mathcal{A}$, then
\begin{multline}
\dfrac{\varphi^{-1}(p_{1})-\varphi^{-1}(q_{1})}{(\varphi^{-1})'(p_{1})}+\dfrac{\varphi^{-1}(p_{2})-\varphi^{-1}(q_{2})}{(\varphi^{-1})'(p_{2})}\\
\geq\dfrac{\varphi^{-1}(p_{1}+p_{2})-\varphi^{-1}(q_{1}+q_{2})}{(\varphi^{-1})'(p_{1}+p_{2})}\label{eq:ineq_partition_equiv}
\end{multline}
is satisfied for all $p_{1},p_{2}$ and $q_{1},q_{2}$ in $(0,1)$
such that the sums $p_{1}+p_{2}$ and $q_{1}+q_{2}$ are in $(0,1)$.
Let us fix $p_{1},p_{2}\in(0,1)$. We rewrite (\ref{eq:ineq_partition_equiv})
as
\begin{multline*}
\dfrac{\varphi^{-1}(p_{1})}{(\varphi^{-1})'(p_{1})}+\dfrac{\varphi^{-1}(p_{2})}{(\varphi^{-1})'(p_{2})}-\dfrac{\varphi^{-1}(p_{1}+p_{2})}{(\varphi^{-1})'(p_{1}+p_{2})}\\
\geq\dfrac{\varphi^{-1}(q_{1})}{(\varphi^{-1})'(p_{1})}+\dfrac{\varphi^{-1}(q_{2})}{(\varphi^{-1})'(p_{2})}-\dfrac{\varphi^{-1}(q_{1}+q_{2})}{(\varphi^{-1})'(p_{1}+p_{2})},
\end{multline*}
which is satisfied if and only if the function
\[
F(q_{1},q_{2})=\dfrac{\varphi^{-1}(q_{1})}{(\varphi^{-1})'(p_{1})}+\dfrac{\varphi^{-1}(q_{2})}{(\varphi^{-1})'(p_{2})}-\dfrac{\varphi^{-1}(q_{1}+q_{2})}{(\varphi^{-1})'(p_{1}+p_{2})}
\]
attains a global maximum at $(q_{1},q_{2})=(p_{1},p_{2})$. By a simple
calculation, it can be verified that $\nabla F(p_{1},p_{2})=0$. Moreover,
we express the determinant of the Hessian of $F$ at $(p_{1},p_{2})$
as
\begin{align*}
J_{F}(p_{1},p_{2}) & =\dfrac{(\varphi^{-1})''(p_{1})}{(\varphi^{-1})'(p_{1})}\dfrac{(\varphi^{-1})''(p_{2})}{(\varphi^{-1})'(p_{2})}\\
 & \qquad-\dfrac{(\varphi^{-1})''(p_{1}+p_{2})}{(\varphi^{-1})'(p_{1}+p_{2})}\dfrac{(\varphi^{-1})''(p_{2})}{(\varphi^{-1})'(p_{2})}\\
 & \qquad-\dfrac{(\varphi^{-1})''(p_{1})}{(\varphi^{-1})'(p_{1})}\dfrac{(\varphi^{-1})''(p_{1}+p_{2})}{(\varphi^{-1})'(p_{1}+p_{2})}\\
 & =\frac{1}{g(p_{1})}\frac{1}{g(p_{2})}-\frac{1}{g(p_{1}+p_{2})}\biggl[\frac{1}{g(p_{1})}+\frac{1}{g(p_{2})}\biggr].
\end{align*}
Because $J_{F}(p_{1},p_{2})\geq0$, it follows that $g(p_{1}+p_{2})\geq g(p_{1})+g(p_{2})$.
\end{IEEEproof}

\begin{rem}
If $\varphi(x)=\exp(x)$ the function $g=-\dfrac{(\varphi^{-1})'}{(\varphi^{-1})''}$
is the identity function which is additive, therefore superadditive.
\end{rem}

\subsection{Joint convexity}

In this section, we find a sufficient condition for the joint convexity
of $D_{\varphi}(\cdot\mathbin{||}\cdot)$. We also show that $D_{\varphi}(\cdot\mathbin{||}\cdot)$
satisfies the partition inequality, and is jointly convex, if, and
only if, the deformed exponential function is a scaled and translated
version of the Tsallis exponential.

The generalized divergence $D_{\varphi}(\cdot\mathbin{||}\cdot)$
is said to be jointly convex if the inequality
\begin{multline}
D_{\varphi}(\lambda\boldsymbol{p}_{1}+(1-\lambda)\boldsymbol{p}_{2}\mathbin{||}\lambda\boldsymbol{q}_{1}+(1-\lambda)\boldsymbol{q}_{2})\\
\leq\lambda D_{\varphi}(\boldsymbol{p}_{1}\mathbin{||}\boldsymbol{q}_{1})+(1-\lambda)D_{\varphi}(\boldsymbol{p}_{2}\mathbin{||}\boldsymbol{q}_{2})\label{eq:jointly_convex}
\end{multline}
is satisfied for all probability distributions $\boldsymbol{p}_{1},\boldsymbol{p}_{2}$
and $\boldsymbol{q}_{1},\boldsymbol{q}_{2}$ in $\Delta_{n}^{\circ}$,
and each $\lambda\in[0,1]$.

Before we find a sufficient condition for the joint convexity of $D_{\varphi}(\cdot\mathbin{||}\cdot)$,
we show some preliminary results.

\begin{lem}
\label{lem:convexity_equivalence_gh} The function $g=-\dfrac{(\varphi^{-1})'}{(\varphi^{-1})''}$
is (strictly) concave if and only if $h=\dfrac{\varphi'}{\varphi''}$
is (strictly) concave.
\end{lem}
\begin{IEEEproof}
Inserting the expressions $(\varphi^{-1})'=1/\varphi'(\varphi^{-1})$
and
\[
(\varphi^{-1})''=-\frac{\varphi''(\varphi^{-1})\cdot(\varphi^{-1})'}{[\varphi'(\varphi^{-1})]^{2}}
\]
into the definition of $g$, we can write
\[
g=\varphi'(\varphi^{-1})\frac{\varphi'(\varphi^{-1})}{\varphi''(\varphi^{-1})}=\varphi'(\varphi^{-1})h(\varphi^{-1}).
\]
Some calculations show that
\[
g_{+}'=1+h_{+}'(\varphi^{-1}),
\]
where $(\cdot)_{+}'$ denotes the right derivative. By the fact of
$\varphi^{-1}$ is strictly increasing, we conclude that $g_{+}'$
is (strictly) decreasing if and only if $h_{+}'$ is (strictly) decreasing.
As a result, for $g$ to be (strictly) concave, it is necessary and
sufficient that $h$ be (strictly) concave.
\end{IEEEproof}

\begin{lem}
\label{lem:F_alpha} The function $g=-\dfrac{(\varphi^{-1})'}{(\varphi^{-1})''}$
is concave if and only if the mapping
\[
F_{\alpha}(x,y)=\varphi((1-\alpha)\varphi^{-1}(x)+\alpha\varphi^{-1}(y)),\quad(x,y)\in\mathbb{R}^{2},
\]
is concave for each $\alpha\in(0,1)$.
\end{lem}
\begin{IEEEproof}
Let us denote $z_{\alpha}=(1-\alpha)\varphi^{-1}(x)+\alpha\varphi^{-1}(y)$.
Some calculations show that
\begin{align*}
\frac{\partial^{2}F_{\alpha}}{\partial x^{2}}(x,y) & =(1-\alpha)(\varphi^{-1})''(x)\varphi'(z_{\alpha})\\
 & \qquad+[(1-\alpha)(\varphi^{-1})'(x)]^{2}\varphi''(z_{\alpha}),\\
\frac{\partial^{2}F_{\alpha}}{\partial y^{2}}(x,y) & =\alpha(\varphi^{-1})''(y)\varphi'(z_{\alpha})\\
 & \qquad+[\alpha(\varphi^{-1})'(y)]^{2}\varphi''(z_{\alpha}),\\
\intertext{and}\frac{\partial^{2}F_{\alpha}}{\partial x\partial y}(x,y) & =\alpha(1-\alpha)(\varphi^{-1})'(x)(\varphi^{-1})'(y)\varphi''(z_{\alpha}),
\end{align*}
which we use to find the following expression for the determinant
of the Hessian of $F_{\alpha}$ at $(x,y)$:
\begin{multline*}
J_{F_{\alpha}}(x,y)=\alpha(1-\alpha)\varphi'(z_{\alpha})\varphi''(z_{\alpha})(\varphi^{-1})''(x)(\varphi^{-1})''(y)\\
\cdot\biggl\{\frac{\varphi'(z_{\alpha})}{\varphi''(z_{\alpha})}+\alpha\frac{[(\varphi^{-1})'(y)]^{2}}{(\varphi^{-1})''(y)}+(1-\alpha)\frac{[(\varphi^{-1})'(x)]^{2}}{(\varphi^{-1})''(x)}\biggr\}.
\end{multline*}
Denote $h=\varphi'/\varphi''$. Noticing that
\begin{equation}
\frac{[(\varphi^{-1})']^{2}}{(\varphi^{-1})''}=-\frac{\varphi'(\varphi^{-1})}{\varphi''(\varphi^{-1})},\label{eq:useful_expression_h}
\end{equation}
we conclude that $J_{F_{\alpha}}(x,y)\geq0$ is equivalent to $h(z_{\alpha})\geq(1-\alpha)h(\varphi^{-1}(x))+\alpha h(\varphi^{-1}(y))$.

To show that the Hessian of $F_{\alpha}$ is negative semi-definitive,
we have to verify, in addition, that its trace is non-positive. Since
$h$ is concave and non-negative, we have
\begin{equation}
\frac{\varphi'(z_{\alpha})}{\varphi''(z_{\alpha})}-(1-\alpha)\frac{\varphi'(\varphi^{-1}(x))}{\varphi''(\varphi^{-1}(x))}\geq0.\label{eq:ineq_conseq_h_concave}
\end{equation}
If we insert (\ref{eq:useful_expression_h}) into (\ref{eq:ineq_conseq_h_concave}),
and multiply the resulting expression by $(1-\alpha)\varphi''(z_{\alpha})(\varphi^{-1})''(x)\leq0$,
we get
\begin{multline*}
\frac{\partial^{2}F_{\alpha}}{\partial x^{2}}(x,y)=(1-\alpha)(\varphi^{-1})''(x)\varphi'(z_{\alpha})\\
+[(1-\alpha)(\varphi^{-1})'(x)]^{2}\varphi''(z_{\alpha})\leq0.
\end{multline*}
Analogously, we also have $(\partial^{2}F_{\alpha}/\partial x^{2})(x,y)\leq0$.
Consequently, the Hessian of $F_{\alpha}$ has a negative trace.

From Lemma~\ref{lem:convexity_equivalence_gh}, it follows that $g$
is concave if and only if $F_{\alpha}$ is concave for each $\alpha\in(0,1)$.
\end{IEEEproof}

\begin{prop}
\label{prop:sufficiency_joint_convexity} If the function $g=-\dfrac{(\varphi^{-1})'}{(\varphi^{-1})''}$
is concave, then the divergence $D_{\varphi}(\cdot\mathbin{||}\cdot)$
is jointly convex.
\end{prop}
\begin{IEEEproof}
According to Lemma~\ref{lem:F_alpha}, the mapping
\[
F_{\alpha}(x,y)=\varphi((1-\alpha)\varphi^{-1}(x)+\alpha\varphi^{-1}(y)),\quad(x,y)\in\mathbb{R}^{2},
\]
is concave for each $\alpha\in(0,1)$. Fixed an arbitrary $\boldsymbol{p}_{j}=(p_{ji})$
and $\boldsymbol{q}_{j}=(q_{ji})$ in $\Delta_{n}^{\circ}$ for $j=0,1$,
define
\begin{align*}
\boldsymbol{p}_{\lambda} & =(1-\lambda)\boldsymbol{p}_{0}+\lambda\boldsymbol{p}_{1},\\
\boldsymbol{q}_{\lambda} & =(1-\lambda)\boldsymbol{q}_{0}+\lambda\boldsymbol{q}_{1},
\end{align*}
for each $\lambda\in(0,1)$. Hence we can write
\begin{multline}
\frac{1}{1-\alpha}\sum_{i=1}^{n}[p_{\lambda i}-F_{\alpha}(q_{\lambda i},p_{\lambda i})]\\
\leq(1-\lambda)\frac{1}{1-\alpha}\sum_{i=1}^{n}[p_{0i}-F_{\alpha}(q_{0i},p_{0i})]\\
+\lambda\frac{1}{1-\alpha}\sum_{i=1}^{n}[p_{1i}-F_{\alpha}(q_{1i},p_{1i})].\label{eq:F_alpha_p_q}
\end{multline}
Using L'H\^{o}pital's rule in the limit below, we obtain
\begin{align*}
\lim_{\alpha\uparrow1}\frac{y-F_{\alpha}(x,y)}{1-\alpha} & =\lim_{\alpha\uparrow1}\frac{y-\varphi((1-\alpha)\varphi^{-1}(x)+\alpha\varphi^{-1}(y))}{1-\alpha}\\
 & =\varphi'(\varphi^{-1}(y))[-\varphi^{-1}(x)+\varphi^{-1}(y)]\\
 & =\frac{\varphi^{-1}(y)-\varphi^{-1}(x)}{(\varphi^{-1})'(y)}.
\end{align*}
Thus, in the limit $\alpha\uparrow1$, expression (\ref{eq:F_alpha_p_q})
becomes
\[
D_{\varphi}(\boldsymbol{p}_{\lambda}\mathbin{||}\boldsymbol{q}_{\lambda})\leq(1-\lambda)D_{\varphi}(\boldsymbol{p}_{0}\mathbin{||}\boldsymbol{q}_{0})+\lambda D_{\varphi}(\boldsymbol{p}_{1}\mathbin{||}\boldsymbol{q}_{1}),
\]
which is the desired result.
\end{IEEEproof}

The next result is a partial converse of Proposition~\ref{prop:sufficiency_joint_convexity}.

\begin{lem}
\label{lem:joint_convexity} If the divergence $D_{\varphi}(\cdot\mathbin{||}\cdot)$
is jointly convex for some $n\geq3$, then the function $g=-\dfrac{(\varphi^{-1})'}{(\varphi^{-1})''}$
satisfies the inequality
\begin{equation}
g\Bigl(\frac{x+y}{2}\Bigr)\geq\frac{g(x)+g(y)}{2},\label{eq:almost_concave}
\end{equation}
for all $x,y\in(0,1)$ such that $x+y\in(0,1)$.
\end{lem}
\begin{IEEEproof}
If $\boldsymbol{p}_{1}=(p_{1i})$, $\boldsymbol{p}_{2}=(p_{2i})$
and $\boldsymbol{q}_{1}=(q_{1i})$, $\boldsymbol{q}_{2}=(q_{2i})$
then inequality (\ref{eq:jointly_convex}) is equivalent to
\begin{multline}
\sum_{i=1}^{n}\Bigl[\lambda\dfrac{\varphi^{-1}(p_{1i})}{(\varphi^{-1})'(p_{1i})}+(1-\lambda)\dfrac{\varphi^{-1}(p_{2i})}{(\varphi^{-1})'(p_{2i})}\\
-\dfrac{\varphi^{-1}(\lambda p_{1i}+(1-\lambda)p_{2i})}{(\varphi^{-1})'(\lambda p_{1i}+(1-\lambda)p_{2i})}\Bigr]\\
\geq\sum_{i=1}^{n}\Bigl[\lambda\dfrac{\varphi^{-1}(q_{1i})}{(\varphi^{-1})'(p_{1i})}+(1-\lambda)\dfrac{\varphi^{-1}(q_{2i})}{(\varphi^{-1})'(p_{2i})}\\
-\dfrac{\varphi^{-1}(\lambda q_{1i}+(1-\lambda)q_{2i})}{(\varphi^{-1})'(\lambda p_{1i}+(1-\lambda)p_{2i})}\Bigr].\label{eq:jointly_convex_equiv}
\end{multline}
For the fixed probability distributions
\begin{align}
\boldsymbol{p}_{1} & =(p_{1},p_{2},p,p_{13},\dots,p_{1n}),\label{eq:p_1}\\
\boldsymbol{p}_{2} & =(p_{2},p_{1},p,p_{23},\dots,p_{2n}),
\end{align}
in $\Delta_{n}^{\circ}$, we consider
\begin{align}
\boldsymbol{q}_{1} & =(p_{1}+x,p_{2}+y,p-x-y,p_{13},\dots,p_{1n}),\\
\boldsymbol{q}_{2} & =(p_{2}+y,p_{1}+x,p-x-y,p_{23},\dots,p_{2n}),\label{eq:q_2}
\end{align}
where $x$ and $y$ are taken so that $\boldsymbol{q}_{1}$ and $\boldsymbol{q}_{2}$
are in $\Delta_{n}^{\circ}$. Inserting these probability distributions
into (\ref{eq:jointly_convex_equiv}) with $\lambda=1/2$, we can
infer that the function
\begin{multline*}
F(x,y)=\frac{1}{2}\dfrac{\varphi^{-1}(p_{1}+x)}{(\varphi^{-1})'(p_{1})}+\frac{1}{2}\dfrac{\varphi^{-1}(p_{2}+y)}{(\varphi^{-1})'(p_{2})}\\
-\dfrac{\varphi^{-1}(\frac{1}{2}(p_{1}+x)+\frac{1}{2}(p_{2}+y))}{(\varphi^{-1})'(\frac{1}{2}p_{1}+\frac{1}{2}p_{2})}+\frac{1}{2}\dfrac{\varphi^{-1}(p_{2}+y)}{(\varphi^{-1})'(p_{2})}\\
+\frac{1}{2}\dfrac{\varphi^{-1}(p_{1}+x)}{(\varphi^{-1})'(p_{1})}-\dfrac{\varphi^{-1}(\frac{1}{2}(p_{2}+y)+\frac{1}{2}(p_{1}+x))}{(\varphi^{-1})'(\frac{1}{2}p_{2}+\frac{1}{2}p_{1})}
\end{multline*}
attains a global maximum at $(x,y)=(0,0)$. Further, we can also write
\begin{multline*}
J_{F}(0,0)=\Bigl[\frac{1}{g(p_{1})}-\frac{1}{2}\frac{1}{g(\frac{1}{2}p_{1}+\frac{1}{2}p_{2})}\Bigr]\\
\cdot\Bigl[\frac{1}{g(p_{2})}-\frac{1}{2}\frac{1}{g(\frac{1}{2}p_{1}+\frac{1}{2}p_{2})}\Bigr]-\Bigl[\frac{1}{2}\frac{1}{g(\frac{1}{2}p_{1}+\frac{1}{2}p_{2})}\Bigr]^{2}\\
=\frac{1}{g(p_{1})}\frac{1}{g(p_{2})}-\frac{1}{g(\frac{1}{2}p_{1}+\frac{1}{2}p_{2})}\Bigl[\frac{1}{2}\frac{1}{g(p_{2})}+\frac{1}{2}\frac{1}{g(p_{1})}\Bigr],
\end{multline*}
where $J_{F}(0,0)$ is the determinant of the Hessian of $F$ at $(0,0)$.
Since $F(x,y)$ attains a maximum at $(0,0)$, inequality $J_{F}(0,0)\geq0$
implies $g(\frac{1}{2}p_{1}+\frac{1}{2}p_{2})\geq\frac{1}{2}g(p_{1})+\frac{1}{2}g(p_{2})$.
\end{IEEEproof}

\begin{prop}
Assume that $n\geq3$. Then the generalized divergence $D_{\varphi}(\cdot\mathbin{||}\cdot)$
satisfies the partition inequality, and is jointly convex, if, and
only if,
\[
\varphi^{-1}(x)=b\ln_{q}(x)-a,\qquad\text{for }x\in(0,1),
\]
for some $q>0$ and $b>0$, $a\in\mathbb{R}$.
\end{prop}
\begin{IEEEproof}
Clearly, inequalities (\ref{eq:almost_superadditive}) and (\ref{eq:almost_concave})
are satisfied for all $x,y\in(0,1]$. Therefore, the function $g(x)$
is superadditive and concave for $x\in(0,1/2)$. It is easy to verify
that $g(0+)=0$. To see this, we apply the limit $x\downarrow0$ in
$0\leq g(x)\leq g(x+y)-g(y)$, and use the continuity of $g$ at $y$.
In addition, because $g(x)$ is concave with $g(0+)=0$, the function
$g(x)$ is also subadditive for $x\in(0,1/2)$. Making $y\downarrow0$
in $g(\lambda x+(1-\lambda)y)\geq\lambda g(x)+(1-\lambda)g(y)$, we
obtain that $g(\lambda x)\geq\lambda g(x)$ for $\lambda\in[0,1]$.
From the inequalities $g(x)\geq\frac{x}{x+y}g(x+y)$ and $g(y)\geq\frac{y}{x+y}g(x+y)$,
it follows that $g(x)+g(y)\geq g(x+y)$. Hence we conclude that $g(x)$
is additive for $x\in(0,1/2)$.

By \cite[Theorem~13.5.2]{Kuczma2009}, there exists $q>0$ such that
$g(x)=x/q$ for $x\in(0,1/2)$. Using (\ref{eq:almost_superadditive}),
and letting $y\downarrow0$ in (\ref{eq:almost_concave}), we get
\[
g(x)\geq g\Bigl(\frac{x}{2}\Bigr)+g\Bigl(\frac{x}{2}\Bigr),\qquad\text{and}\qquad g\Bigl(\frac{x}{2}\Bigr)\geq\frac{g(x)}{2},
\]
which imply $g(x)=2g(x/2)$ for all $x\in(0,1)$. Hence, expression
$g(x)=x/q$ is also verified for $x\in(0,1)$. Solving
\[
g(x)=-\frac{(\varphi^{-1})'(x)}{(\varphi^{-1})''(x)}=\frac{x}{q}
\]
with respect to $\varphi^{-1}(x)$, we find $b>0$ and $a\in\mathbb{R}$
such that
\begin{align*}
\varphi^{-1}(x) & =b\frac{x^{1-q}}{1-q}-a\\
 & =b\ln_{q}(x)-a,\qquad\text{for }q\neq1,
\end{align*}
and
\[
\varphi^{-1}(x)=b\ln(x)-a,\qquad\text{for }q=1,
\]
for every $x\in(0,1)$.

The converse direction follows from Propositions~\ref{prop:ineq_partition}
and~\ref{prop:sufficiency_joint_convexity}.
\end{IEEEproof}

\subsection{Pinsker's inequality}

Pinsker's inequality relates the divergence with the $\ell_{1}$-distance.
This inequality implies that convergence in divergence is stronger
than convergence in the $\ell_{1}$-distance For the KL divergence,
Pinsker's inequality is given by
\begin{equation}
D_{\mathrm{KL}}(\boldsymbol{p}\mathbin{||}\boldsymbol{q})\geq\frac{1}{2}\Vert\boldsymbol{p}-\boldsymbol{q}\Vert_{1}^{2},\label{eq:Pinsker_inequality}
\end{equation}
where $\Vert\boldsymbol{p}-\boldsymbol{q}\Vert_{1}=\sum_{i=1}^{n}\mathbin{|}p_{i}-q_{i}\mathbin{|}$
is the $\ell_{1}$-distance between probability distributions $\boldsymbol{p}=(p_{i})$
and $\boldsymbol{q}=(q_{i})$ in $\Delta_{n}^{\circ}$.

The next result shows Pinsker's inequality for the generalized divergence.

\begin{thm}[Pinsker's Inequality]
 Suppose that the partition inequality (\ref{eq:ineq_partition})
holds. In addition, assume that
\begin{equation}
c=\inf_{0<p<q<1}\frac{1}{8}\frac{1}{q-p}\biggl[-\frac{(\varphi^{-1})'(q)}{(\varphi^{-1})'(p)}+\frac{(\varphi^{-1})'(1-q)}{(\varphi^{-1})'(1-p)}\biggr]>0.\label{eq:c}
\end{equation}
Then, for any probability distributions $\boldsymbol{p}=(p_{i})$
and $\boldsymbol{q}=(q_{i})$ in $\Delta_{n}^{\circ}$, the generalized
divergence satisfies the inequality
\begin{equation}
D_{\varphi}(\boldsymbol{p}\mathbin{||}\boldsymbol{q})\geq c\Vert\boldsymbol{p}-\boldsymbol{q}\Vert_{1}^{2}.\label{eq:ineq_Pinsker}
\end{equation}
\end{thm}
\begin{IEEEproof}
Let $\mathcal{A}=\{A_{1},A_{2}\}$ be a partition of $I_{n}$, where
$A_{1}=\{i:\;p_{i}\geq q_{i}\}$ and $A_{2}=\{i:\;p_{i}<q_{i}\}$.
Hence we can write
\begin{align*}
\Vert\boldsymbol{p}-\boldsymbol{q}\Vert_{1} & =\sum_{i=1}^{n}|p_{i}-q_{i}|\\
 & =\sum_{i\in A_{1}}(p_{i}-q_{i})+\sum_{i\in A_{2}}(q_{i}-p_{i})\\
 & =(p_{1}^{\mathcal{A}}-q_{1}^{\mathcal{A}})+(q_{2}^{\mathcal{A}}-p_{2}^{\mathcal{A}})\\
 & =\Vert\boldsymbol{p}^{\mathcal{A}}-\boldsymbol{q}^{\mathcal{A}}\Vert_{1}.
\end{align*}
By the partition inequality
\[
D_{\varphi}(\boldsymbol{p}\mathbin{||}\boldsymbol{q})\geq D_{\varphi}(\boldsymbol{p}^{\mathcal{A}}\mathbin{||}\boldsymbol{q}^{\mathcal{A}}),
\]
we see that it suffices to show
\begin{equation}
D_{\varphi}(\boldsymbol{p}^{\mathcal{A}}\mathbin{||}\boldsymbol{q}^{\mathcal{A}})\geq c\Vert\boldsymbol{p}^{\mathcal{A}}-\boldsymbol{q}^{\mathcal{A}}\Vert_{1}^{2}.\label{eq:ineq_Pinsker_partition}
\end{equation}
Let us denote $p_{1}^{\mathcal{A}}=p$ and $q_{1}^{\mathcal{A}}=q$.
Then inequality (\ref{eq:ineq_Pinsker_partition}) can be rewritten
as
\[
\frac{\varphi^{-1}(p)-\varphi^{-1}(q)}{(\varphi^{-1})'(p)}+\frac{\varphi^{-1}(1-p)-\varphi^{-1}(1-q)}{(\varphi^{-1})'(1-p)}\geq4c(p-q)^{2},
\]
since $\Vert\boldsymbol{p}^{\mathcal{A}}-\boldsymbol{q}^{\mathcal{A}}\Vert_{1}=2(p-q)$.
For a fixed $p\in(0,1)$, we define the function
\begin{multline*}
F(q)=\frac{\varphi^{-1}(p)-\varphi^{-1}(q)}{(\varphi^{-1})'(p)}\\
+\frac{\varphi^{-1}(1-p)-\varphi^{-1}(1-q)}{(\varphi^{-1})'(1-p)}-4c(p-q)^{2},
\end{multline*}
for $q\in(0,1)$. By the symmetry of the terms $p$ and $q$ in (\ref{eq:c}),
it is clear that
\[
c=\inf_{0<q<p<1}\frac{1}{8}\frac{1}{q-p}\biggl[-\frac{(\varphi^{-1})'(q)}{(\varphi^{-1})'(p)}+\frac{(\varphi^{-1})'(1-q)}{(\varphi^{-1})'(1-p)}\biggr]>0.
\]
As a result, the derivative
\[
F^{\prime}(q)=(q-p)\biggl\{\frac{1}{q-p}\biggl[-\frac{(\varphi^{-1})'(q)}{(\varphi^{-1})'(p)}+\frac{(\varphi^{-1})'(1-q)}{(\varphi^{-1})'(1-p)}\biggr]-8c\biggr\}
\]
is $\geq0$ for $q>p$, and $\leq0$ for $q<p$. We conclude that
$F(q)$ attains a minimum at $q=p$. Therefore,
\[
D_{\varphi}(\boldsymbol{p}^{\mathcal{A}}\mathbin{||}\boldsymbol{q}^{\mathcal{A}})-c\Vert\boldsymbol{p}^{\mathcal{A}}-\boldsymbol{q}^{\mathcal{A}}\Vert_{1}^{2}=F(q)\geq F(p)=0,
\]
and inequality (\ref{eq:ineq_Pinsker}) follows.
\end{IEEEproof}

If we assume $\varphi^{-1}(x)=\log(x)$, then expression (\ref{eq:c})
results in $c=1/2$, which is the constant in Pinsker's inequality
for the KL divergence. For the Tsallis exponential, an easy computation
shows that $c=q/2$ in equation (\ref{eq:c}) with $\varphi^{-1}(x)=\ln_{q}(x)$.
This result is in accordance to the work of Gilardoni \cite{Gilardoni2010},
which investigated the Pinsker's inequality for $f$-divergences.
Gilardoni showed that the $f$-divergence $D_{f}(\boldsymbol{p}\mathbin{||}\boldsymbol{q})=\sum_{i=1}^{n}p_{i}f\bigl(\frac{q_{i}}{p_{i}}\bigr)$
satisfies the inequality $D_{f}(\boldsymbol{p}\mathbin{||}\boldsymbol{q})\geq\frac{f^{\prime\prime}(1)}{2}\Vert\boldsymbol{p}-\boldsymbol{q}\Vert_{1}^{2}$,
supposing that $f$ is convex and three times differentiable at $x=1$
with $f^{\prime\prime}(1)>0$. Tsallis relative entropy is an $f$-divergence
with $f(x)=-\ln_{q}(x)$. In this case, we have $f''(1)=q$.

\section{\label{sec:conclusions}Conclusions}

In this work, we found necessary and sufficient conditions for the
generalized divergence $D_{\varphi}(\cdot\mathbin{||}\cdot)$ to satisfy
the partition inequality. We also showed a condition that implies
the joint convexity of $D_{\varphi}(\cdot\mathbin{||}\cdot)$. It
was proved that, for the generalized divergence $D_{\varphi}(\cdot\mathbin{||}\cdot)$
to coincide with the Tsallis relative entropy $D_{q}(\cdot\mathbin{||}\cdot)$,
it is necessary and sufficient that $D_{\varphi}(\cdot\mathbin{||}\cdot)$
satisfy the partition inequality, and be jointly convex. As an application
of partition inequality, a criterion for the Pinsker's inequality
was found. We also constructed a family of probability distributions
associated with the generalized divergence.

This work can be extended in many aspects. The data processing inequality
was not proved. Comparisons between generalized divergences, as investigated
in \cite{Harremoees2011} for $f$-divergences, have the potential
of being a prosperous topic of research. In \cite{Souza2016}, a generalization
of R\'{e}nyi divergence was defined in terms of a deformed exponential.
As future work, we aim to investigate the properties of this generalized
R\'{e}nyi divergence.

\section*{Acknowledgment}

The authors would like to thank CNPq (Procs.\ 408609/2016-8 and 309472/2017-2) and Coordena\c{c}\~{a}o de Aperfei\c{c}oamento de Pessoal de N\'{\i}vel Superior - Brazil (CAPES) - Finance Code 001 for partial funding of this research. We would also like to thank
Sueli I.R.\ Costa for the valuable contributions to this work.

%\bibliographystyle{IEEEtran}
%\bibliography{IEEEabrv,refs_gen_entropy}

\begin{thebibliography}{1}
	\providecommand{\url}[1]{#1}
	\csname url@samestyle\endcsname
	\providecommand{\newblock}{\relax}
	\providecommand{\bibinfo}[2]{#2}
	\providecommand{\BIBentrySTDinterwordspacing}{\spaceskip=0pt\relax}
	\providecommand{\BIBentryALTinterwordstretchfactor}{4}
	\providecommand{\BIBentryALTinterwordspacing}{\spaceskip=\fontdimen2\font plus
		\BIBentryALTinterwordstretchfactor\fontdimen3\font minus
		\fontdimen4\font\relax}
	\providecommand{\BIBforeignlanguage}[2]{{%
			\expandafter\ifx\csname l@#1\endcsname\relax
			\typeout{** WARNING: IEEEtran.bst: No hyphenation pattern has been}%
			\typeout{** loaded for the language `#1'. Using the pattern for}%
			\typeout{** the default language instead.}%
			\else
			\language=\csname l@#1\endcsname
			\fi
			#2}}
	\providecommand{\BIBdecl}{\relax}
	\BIBdecl
	
	\bibitem{Cover2006}
	T.~M. Cover and J.~A. Thomas, \emph{Elements of information theory},
	2nd~ed.\hskip 1em plus 0.5em minus 0.4em\relax Wiley-Interscience [John Wiley
	\& Sons], Hoboken, NJ, 2006.
	
	\bibitem{Hastie2009}
	T.~Hastie, R.~Tibshirani, and J.~Friedman, \emph{The elements of statistical
		learning}, 2nd~ed., ser. Springer Series in Statistics.\hskip 1em plus 0.5em
	minus 0.4em\relax Springer, New York, 2009, data mining, inference, and
	prediction.
	
	\bibitem{Principe2010}
	J.~C. Principe, \emph{Information theoretic learning}, ser. Information Science
	and Statistics.\hskip 1em plus 0.5em minus 0.4em\relax Springer, New York,
	2010, {R}\'{e}nyi's entropy and kernel perspectives.
	
	\bibitem{Konishi2008}
	S.~Konishi and G.~Kitagawa, \emph{Information criteria and statistical
		modeling}, ser. Springer Series in Statistics.\hskip 1em plus 0.5em minus
	0.4em\relax Springer, New York, 2008.
	
	\bibitem{Kullback1951}
	\BIBentryALTinterwordspacing
	S.~Kullback and R.~A. Leibler, ``On information and sufficiency,'' \emph{Ann.
		Math. Statistics}, vol.~22, pp. 79--86, 1951. [Online]. Available:
	\url{https://doi.org/10.1214/aoms/1177729694}
	\BIBentrySTDinterwordspacing
	
	\bibitem{Borland1998}
	\BIBentryALTinterwordspacing
	L.~Borland, A.~R. Plastino, and C.~Tsallis, ``Information gain within
	nonextensive thermostatistics,'' \emph{J. Math. Phys.}, vol.~39, no.~12, pp.
	6490--6501, 1998. [Online]. Available: \url{https://doi.org/10.1063/1.532660}
	\BIBentrySTDinterwordspacing
	
	\bibitem{Borland1999}
	\BIBentryALTinterwordspacing
	------, ``Erratum: ``{I}nformation gain within generalized [non-extensive]
	thermostatistics'','' \emph{J. Math. Phys.}, vol.~40, no.~4, p. 2196, 1999.
	[Online]. Available: \url{https://doi.org/10.1063/1.533119}
	\BIBentrySTDinterwordspacing
	
	\bibitem{Erven2014}
	\BIBentryALTinterwordspacing
	T.~van Erven and P.~Harremo\"{e}s, ``R\'{e}nyi divergence and
	{K}ullback-{L}eibler divergence,'' \emph{IEEE Trans. Inform. Theory},
	vol.~60, no.~7, pp. 3797--3820, 2014. [Online]. Available:
	\url{https://doi.org/10.1109/TIT.2014.2320500}
	\BIBentrySTDinterwordspacing
	
	\bibitem{Furuichi2004}
	S.~Furuichi, K.~Yanagi, and K.~Kuriyama, ``Fundamental properties of {T}sallis
	relative entropy,'' \emph{J. Math. Phys.}, vol.~45, no.~12, pp. 4868--4877,
	2004.
	
	\bibitem{Furuichi2005}
	\BIBentryALTinterwordspacing
	S.~Furuichi, ``On uniqueness theorems for {T}sallis entropy and {T}sallis
	relative entropy,'' \emph{IEEE Trans. Inform. Theory}, vol.~51, no.~10, pp.
	3638--3645, 2005. [Online]. Available:
	\url{https://doi.org/10.1109/TIT.2005.855606}
	\BIBentrySTDinterwordspacing
	
	\bibitem{Zhang2004}
	J.~Zhang, ``Divergence function, duality, and convex analysis,'' \emph{Neural
		Comput.}, vol.~16, no.~1, pp. 159--195, Jan. 2004.
	
	\bibitem{Broniatowski2019}
	M.~Broniatowski and W.~Stummer, ``Some universal insights on divergences for
	statistics, machine learning and artificial intelligence,'' in
	\emph{Geometric structures of information}, ser. Signals Commun.
	Technol.\hskip 1em plus 0.5em minus 0.4em\relax Springer, Cham, 2019, pp.
	149--211.
	
	\bibitem{Gilardoni2010}
	G.~L. Gilardoni, ``On {P}insker's and {V}ajda's type inequalities for
	{C}sisz\'ar's {$f$}-divergences,'' \emph{IEEE Trans. Inform. Theory},
	vol.~56, no.~11, pp. 5377--5386, 2010.
	
	\bibitem{Sason2016}
	\BIBentryALTinterwordspacing
	I.~Sason and S.~Verd\'{u}, ``{$f$}-divergence inequalities,'' \emph{IEEE Trans.
		Inform. Theory}, vol.~62, no.~11, pp. 5973--6006, 2016. [Online]. Available:
	\url{https://doi.org/10.1109/TIT.2016.2603151}
	\BIBentrySTDinterwordspacing
	
	\bibitem{Bhatia2007}
	R.~Bhatia, \emph{Positive definite matrices}, ser. Princeton Series in Applied
	Mathematics.\hskip 1em plus 0.5em minus 0.4em\relax Princeton University
	Press, Princeton, NJ, 2007.
	
	\bibitem{Hardy1988}
	G.~H. Hardy, J.~E. Littlewood, and G.~P\'{o}lya, \emph{Inequalities}, ser.
	Cambridge Mathematical Library.\hskip 1em plus 0.5em minus 0.4em\relax
	Cambridge University Press, Cambridge, 1988, reprint of the 1952 edition.
	
	\bibitem{Matkowski1990}
	\BIBentryALTinterwordspacing
	J.~Matkowski, ``The converse of the {M}inkowski's inequality theorem and its
	generalization,'' \emph{Proc. Amer. Math. Soc.}, vol. 109, no.~3, pp.
	663--675, 1990. [Online]. Available: \url{https://doi.org/10.2307/2048205}
	\BIBentrySTDinterwordspacing
	
	\bibitem{Kuczma2009}
	M.~Kuczma, \emph{An introduction to the theory of functional equations and
		inequalities}, 2nd~ed.\hskip 1em plus 0.5em minus 0.4em\relax Birkh\"auser
	Verlag, Basel, 2009.
	
	\bibitem{Harremoees2011}
	\BIBentryALTinterwordspacing
	P.~Harremo\"{e}s and I.~Vajda, ``On pairs of {$f$}-divergences and their joint
	range,'' \emph{IEEE Trans. Inform. Theory}, vol.~57, no.~6, pp. 3230--3235,
	2011. [Online]. Available: \url{https://doi.org/10.1109/TIT.2011.2137353}
	\BIBentrySTDinterwordspacing
	
	\bibitem{Souza2016}
	\BIBentryALTinterwordspacing
	D.~C. de~Souza, R.~F. Vigelis, and C.~C. Cavalcante, ``Geometry induced by a
	generalization of {R}\'{e}nyi divergence,'' \emph{Entropy}, vol.~18, no.~11,
	pp. Paper No. 407, 16, 2016. [Online]. Available:
	\url{https://doi.org/10.3390/e18110407}
	\BIBentrySTDinterwordspacing
	
\end{thebibliography}

\def\cprime{$'$}

\begin{IEEEbiographynophoto}{Rui F.\ Vigelis}
received the B.Sc degree in Electrical Engineering from the Federal University of Cear\'{a}, Brazil, in 2005, and the M.Sc and Ph.D.\ degrees in Teleinformatics Engineering from the Federal University of Cear\'{a}, Brazil, in 2006 and 2011, respectively. Since 2012, he is an Assistant Professor at the Federal University of Cear\'{a}, campus Sobral. His primary research interests are in the analysis of non-standard function spaces (e.g., Musielak--Orlicz spaces), non-parametric information geometry, and measures of information.
\end{IEEEbiographynophoto}

\begin{IEEEbiographynophoto}{Luiza H.F.\ Andrade}
received the B.Sc degree in Mathematics from the Cear\'{a} State University, in 2002, the M.Sc degree in Mathematics and the Ph.D.\ degree in Teleinformatics Engineering, both from the Federal University of Cear\'{a}, in 2007 and 2018, respectively. She has been developing research in information geometry and information theory.
\end{IEEEbiographynophoto}

\begin{IEEEbiographynophoto}{Charles C.\ Cavalcante}
(S'98 - M'04 - SM'11) received the B.Sc and M.Sc in Electrical Engineering from the Federal University of Cear\'{a} (UFC), Brazil, in 1999 and 2001, respectively, and the Ph.D.\ degree from the University of Campinas (UNICAMP), Brazil, in 2004. He has held a grant for Scientific and Technological Development from 2004 to 2007 and since March 2009 he has a grant of Scientific Research Productivity both from the Brazilian Research Council (CNPq). He is now an Associate Professor at Teleinformatics Engineering Department of UFC holding the Statistical Signal Processing chair. From August 2014 to July 2015 he was a Visiting Assistant Professor at the Department of Computer Science and Electrical Engineering (CSEE) from University of Maryland, Baltimore County (UMBC) in the United States. He has been working on signal processing strategies for communications where he has several papers published in journal and conferences, has authored three international patents and he has worked on several funded research projects on the signal processing and wireless communications areas. He is also a co-author of the book Unsupervised Signal Processing: Channel Equalization and Source Separation and co-editor of the book Signals and Images: Advances and Results in Speech, Estimation, Compression, Recognition, Filtering, and Processing, both published by CRC Press. He is a researcher of the Wireless Telecommunications Research Group (GTEL) where he leads research on signal processing and wireless communications. Dr. Cavalcante is a Senior Member of the IEEE and Senior Member of the Brazilian Telecommunications Society (SBrT) for the term 2018-2020. Since March 2018 he is the President of the Brazilian Telecommunication Society (SBrT) and has just been elected to the IEEE Signal Processing Society Board of Governors in the capacity of Regional Director-at-Large for Regions 7 \& 9 for the term 2020-2021. His main research interests are in signal processing for communications, statistical signal processing and information geometry.
\end{IEEEbiographynophoto}

\end{document}